\documentclass[11pt]{article}
\usepackage{amsmath,amsfonts,amsthm,amssymb,amscd}
 \begin{document}
\author {{\bf A. Borowiec$^{*}$ and M. Francaviglia$^{**}$}}
\title{{\bf Three-Dimensional Chern-Simons and BF Theories}}
\maketitle
\date{ $\ $ $\ $ $\ $ $\ $ $\ $ $\ $ $\ $ $\ $ $\ $ $\ $ $\ $
AIP Conf.Proc.751:165-167,2005  \\
$^{\ \,*\ }$Institute for Theoretical Physics, University of Wroc{\l}aw, Poland \\
$^{**}$ Dipartimento di Matematica, Universit\`a di Torino, Italy}
 \begin{abstract}
Our aim in this note is to clarify a relationship between covariant
Chern-Simons 3-dimensional theory and Schwartz type topological
field theory known also as BF theory.
\end{abstract}
\maketitle


\section{Introduction.}

Chern-Simons theory gives an interesting example of topological
field theory. Its Lagrangian 3-form lives on a principal $G$-bundle
and after pulling back to space-time (base) manifold provides, in
general, a family of local, non-covariant 
Lagrangian densities
\cite{BFF2,BFFP}. \footnote{However, the corresponding
Euler-Lagrange equations of motion have well-defined global
meaning.} Because of this, it is also more difficult to analyze, in
this case, N\"oether conserved quantities \cite{BFF1}. A more standard
approach to the problem of symmetries and conservation laws has been
applied in the so called covariant formalism \cite{BFF2}. It
exploits the transgression 3-form  as a global and covariant
Chern-Simons Lagrangian with two dynamical gauge fields. This
formalism has been used for the calculation of N\"other currents and
their identically vanishing parts - superpotentials. Augmented
variational principle and relative conservation laws have been
recently proposed in \cite{FFF}. Our aim in the present note, 
which can be viewed as an appendix to \cite{BFF2}, is to explain a 
link between covariant Chern-Simons theory and the so called BF
theories \cite{BBRT,CCFM,Hor}.

\section{Change of variables.}

Let us consider a principal bundle $P(M,G)$ over a three-dimensional
manifold $M$ with a (semisimple) structure group $G$. Let
$\omega_{i}$ \ ($i=0,1$)  be two principal connection 1-forms with the
corresponding curvature 2-forms
\begin{eqnarray}\label{a1}
\Omega_{i}=d\omega_{i}+\omega_{i}^{2}=d\omega_{i}+\frac{1}{2}[\omega
_{i},\omega_{i}] .\end{eqnarray}
Denote by $\alpha=\omega_{1}-\omega_{0}$, a  tensorial 1-form.

The transgression 3-form is given by the well known formula
\begin{eqnarray}\label{a2}
Q\left( \omega_{1},\omega_{0}\right) = -Q\left(\omega_{0},\omega_{1}\right) =tr\left(2\Omega_{0}\wedge\alpha+D_{0}\alpha\wedge\alpha+\frac{2}{3}%
\alpha^{3}\right)\nonumber\\
=tr\left(  2\Omega_{1}\wedge\alpha-D_{1}\alpha\wedge\alpha+\frac{2}{3}\alpha^{3}\right)
\end{eqnarray}
where $D_i\alpha=d\alpha+[\omega_i,\alpha]$ denotes the covariant
derivative of $\alpha$ with respect to the connection $\omega_i$.
Thus $Q(\omega_{1},\omega_{0})$ is a tensorial (covariant) object
which well-defines the corresponding global 3-form on $M$. It
undergoes a non-covariant splitting as a difference of two
Chern-Simons Lagrangians
\begin{eqnarray}\label{a3}
Q\left(  \omega_{1},\omega_{0}\right)  =CS(\omega
_{1})-CS(\omega_{0})+dtr\left(  \omega_{0}\wedge\omega_{1}\right)\end{eqnarray}
where
\begin{eqnarray}\label{a4}
CS(\omega)=tr\left(\Omega\wedge\omega-\frac{1}{3}\omega^{3}\right)\end{eqnarray}
stands for non-covariant Chern-Simons Lagrangian.

Notice that in the case of two connections one has
\begin{eqnarray}\label{a5}
2\Omega_{0}+D_{0}\alpha=2\Omega_{1}-D_{1}\alpha=\nonumber\\
\Omega_{0}+\Omega_{1}%
+\frac{1}{2}\left(  D_{0}\alpha-D_{1}\alpha\right)=
\Omega_{0}+\Omega_{1}-\alpha^{2}
\end{eqnarray}
The last equality entitles us to rewrite
\begin{eqnarray}\label{a6}
Q\left(  \omega_{1},\omega_{0}\right)  =
2tr\left(\bar\Omega\alpha
+\frac{1}{12}\alpha^{3}\right)  \
\end{eqnarray}
where $\bar\omega=\frac{1}{2}\left(  \omega_{1}+\omega_{0}\right)  $ is a new (average)
connection and $\bar\Omega=d\bar\omega+\bar\omega^{2}$. 
Of course, one has $\omega_{1}=\bar\omega+\frac{1}{2}\alpha$ \ , \ \ $\omega_{0}=\bar\omega
-\frac{1}{2}\alpha$.

Thus the Lagrangian $Q\left(  \omega_{1},\omega_{0}\right)  $ can be treated in
three different (but equivalent) ways:
\begin{itemize}
\item with two (flat) connections $\omega_{0},\ \omega_{1}$ as dynamical variables; see (\ref{a3}). In this case $\Omega_{0}=0$  and $\ \Omega_{1}=0$ are  equations of motion. This point of view was presented in \cite{BFF2}.

\item with a (flat) connection $\omega_{1}$ and tensorial 1-form $\alpha$ as dynamical variables; see (\ref{a2}). In this case equations of motion are
$\Omega_{1}=0$ \ and $D_{1}\alpha=\alpha^{2}$.

\item with an "average' (non-flat) connection $\bar\omega=\frac{1}{2}\left(  \omega
_{1}+\omega_{2}\right)$ and tensorial 1-form $\alpha$ as independent dynamical
variables; see (\ref{a6}). In this case
$\bar \Omega=-\frac{1}{4}\alpha^{2}$ \ and \ $\bar D\alpha=0$ are equations of motion.
This is the so called BF theory with a cosmological constant $\Lambda=1$ (see e.g references
\cite{BBRT,CCFM,Mont,Sarda}).
\end{itemize}
\bigskip

More generally, one can define a new connection $\omega_{t}=t\omega_{1}+\left(
1-t\right)  \omega_{0}=\omega_{0}+t\alpha$ as a convex combination
of two connections  with parameter \ $0\leq t\leq1$. The inverse transformation is $\omega_{0}=\omega_t-t\alpha$ \ , $\omega_{1}=\omega_{t}+(1-t)\alpha$. In this case
$$\Omega_{t}=t\Omega_{1}+\left(  1-t\right)  \Omega_{0}-t\left(
1-t\right)  \alpha^{2}\ .$$
Now the equation (\ref{a5}) can be replaced by the more general one
\begin{eqnarray}\label{a7}
2\Omega_{1}-D_{1}\alpha=2\Omega_{0}+D_{0}\alpha=\\\nonumber
=2\Omega_{t}-2t\left(  1-t\right)  \alpha^{2}-\left(  2t-1\right)
D_{t}\alpha
\end{eqnarray}
(notice that $t\omega_{1}+\left(  t-1\right)  \omega_{0}=\left(  2t-1\right)
\omega_{t}+2t\left(  1-t\right)  \alpha$).

In this new variables $(\omega_t,\ \alpha)$ we  obtain
\begin{eqnarray}\label{a8}
Q\left(  \omega_{1},\omega_{0}\right)  =
2tr\left(  \Omega_{t}\wedge\alpha-\left(  t-\frac{1}{2}\right)  D_{t}\alpha
\wedge\alpha+(\frac{1}{3}-t+t^{2})\alpha^{3}\right)
\end{eqnarray}
The corresponding equations of motion are:
\begin{eqnarray}
\Omega_{t}=-t\left(  1-t\right)  \alpha^{2} \ \ \ \ , \ \ \ \ \ \ D_{t}%
\alpha=\left(  2t-1\right)  \alpha^{2}\end{eqnarray}
Thus the choices $t=\frac{1}{2}, 0, 1$ lead to the simplest formulae.

It is interesting to observe that the superpotential related to (infinitesimal) gauge
transformation $\chi$ remains independent of the choice of variables $(\omega_t,\ \alpha)$
(compare formula (19) in \cite{BFF2}), i.e.:
$$
U(\chi)=tr(\alpha\chi) .
$$
Instead, an explicite expression for the superpotential related to (infinitesimal) diffeomorphism transformation driven by a vectorfield $\xi$ does depend on the variables $(\omega_t,\ \alpha)$ and equals
to (compare formulae (23,25) in \cite{BFF2}):
$$
U(\xi)=tr[\alpha(2\omega_t(\xi)+(1-2t)\alpha(\xi))]\ .
$$
   



\bibliographystyle{aipproc}   





\end{document}